\documentclass[12pt,a4paper]{article}

\usepackage[utf8]{inputenc}
\usepackage[T1]{fontenc}
\usepackage{lmodern}
\usepackage{graphicx}
\usepackage{amsmath,amssymb}
\usepackage{hyperref}
\usepackage{natbib}
\usepackage{geometry}
\title{{Jets in Low-Mass Protostars}}
\author{Somnath Dutta$^{1}$\thanks{Corresponding author: sdutta@asiaa.sinica.edu.tw}
}
\date{
  $^{1}$Academia Sinica Institute of Astronomy and Astrophysics, No. 1, Sec. 4,  Roosevelt Rd, Taipei 106319, Taiwan\\[1ex]
}

\begin{document}
\maketitle
\abstract{
%Background:
Jets and outflows are key components of low-mass star formation, regulating accretion and shaping the surrounding molecular clouds. These flows, traced by molecular species at (sub)millimeter wavelengths (e.g., CO, SiO, SO, H$_2$CO, and CH$_3$OH) and by atomic, ionized, and molecular lines in the infrared (e.g., H$_2$, [Fe II], [S I]), originate from protostellar accretion disks deeply embedded within dusty envelopes. Jets play a crucial role in removing angular momentum from the disk, thereby enabling continued mass accretion, while directly preserving a record of the protostar’s outflow history and potentially providing indirect insights into its accretion history. 
%Methods
Recent advances in high-resolution, high-sensitivity observations, particularly with the James Webb Space Telescope (JWST) in the infrared and the Atacama Large Millimeter/submillimeter Array (ALMA) at (sub)millimeter wavelengths, have revolutionized studies of protostellar jets and outflows. These instruments provide complementary views of warm, shock-excited gas and cold molecular component of the jet--outflow system.
%Results:
In this review, we discuss the current status of observational studies that reveal detailed structures, kinematics, and chemical compositions of protostellar jets and outflows. Recent analyses of mass-loss rates, velocities, rotation, molecular abundances, and magnetic fields provide critical insights into jet launching mechanisms, disk evolution, and the potential formation of binary systems and planets.
%Conclusion:
The synergy of JWST’s infrared sensitivity and ALMA’s high-resolution imaging is advancing our understanding of jets and outflows. Future large-scale, high-resolution surveys with these facilities are expected to drive major breakthroughs in outflow research.
}

\section{Introduction}
\begin{figure}%[H]  
    \includegraphics[width=0.6\linewidth]{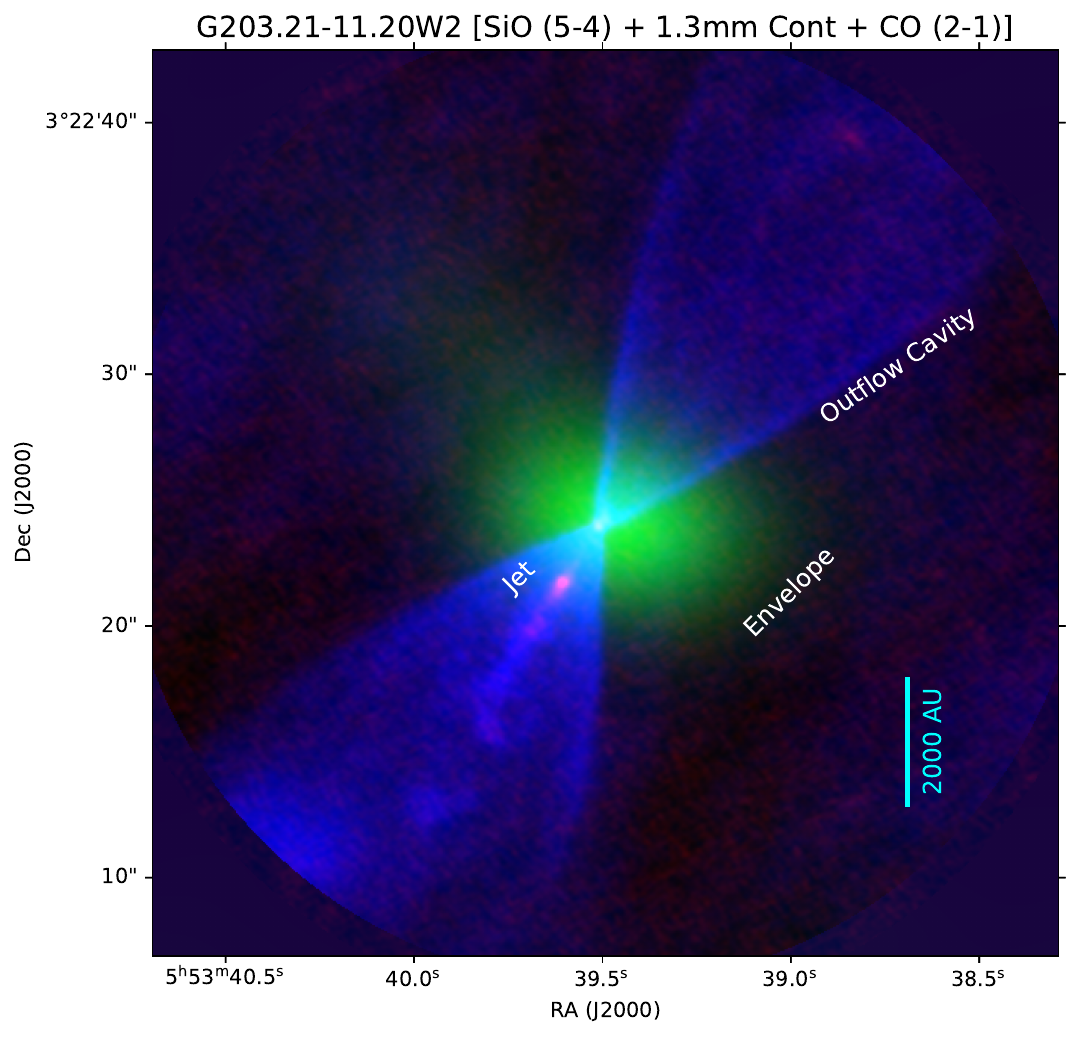}
    \caption{G203.21-11.20W2 protostellar system located in the Orion molecular cloud, observed with ALMA Band 6. The RGB composite shows red: SiO (5–4), green: 1.3 mm continuum, and blue: CO (2–1). A narrow, collimated jet traced by SiO emission is visible along the axis of a wide-angle outflow cavity traced by CO emission. The protostar, embedded within the dense envelope traced by the \mbox{1.3 mm} continuum, lies at the intersection of the bipolar outflow lobes. The SiO and CO maps have an angular resolution of approximately 150 AU, while the 1.3 mm continuum image has a resolution of approximately 2000 AU, with the maximum recoverable scale of $\sim$10,000 AU ($\sim$25$^{\prime\prime}$). This image has been reproduced with data from \citet[][]{2024AJ....167...72D}.}
    \label{fig:ALMA_jet_out_cont}
\end{figure}

Protostellar jets and outflows are ubiquitous phenomena observed during the early stages of star formation. Jets are highly collimated, fast-moving streams of gas ejected from the inner disk regions of young stellar objects (YSOs), often reaching velocities of several hundred kilometers per second. In contrast, winds or outflows generally refer to broader, less collimated flows of gas, typically moving at a few kilometers per second to a few tens of kilometers per second, launched from wider regions of the disk and often entrained by the jets \cite{2001ARA&A..39..403R,2007prpl.conf..245A,2014prpl.conf..451F,2014prpl.conf..387A,2016ARA&A..54..491B,2020A&ARv..28....1L,2021NewAR..9301615R}. Together, jets and outflows play a crucial role in removing excess angular momentum from the protostellar disk, thereby enabling mass accretion onto the central star \cite{2016ARA&A..54..491B,2017NatAs...1E.152L}. An example of a protostellar system, G203.21-11.20W2, is shown in Figure \ref{fig:ALMA_jet_out_cont}. A narrow, collimated jet, traced by SiO emission, is observed along the flow axis, within a wide-angle bipolar outflow cavity delineated by CO emission. The protostar, embedded within a dense envelope traced by 1.3 mm continuum emission, is located at the origin point where the two outflow lobes diverge.

Two primary theoretical frameworks have been proposed for the launching of jets: the X-wind model, where jets originate from a narrow region near the inner edge of the accretion disk at the magnetospheric truncation radius \cite{1994ApJ...429..781S,2000prpl.conf..789S}, and the disk wind model, where winds are launched from a wider range of disk radii via magneto-centrifugal \mbox{forces \cite{2007prpl.conf..277P}.} Both models have been successful in explaining certain observational features, yet debates persist over their relative contributions in different sources. Recent models of both X-winds and magneto-hydrodynamic disk winds indicate that these mechanisms can generate not only highly collimated jets but also wider-angle winds, suggesting that the dichotomy between jets and winds is not always clear-cut \citep{2020ApJ...905..116S}.

Protostellar jets can be observed in molecular, atomic, and ionized forms. Jets are traced across multiwavelengths---from radio, (sub)millimeter, and infrared to \mbox{optical---revealing} their structure and dynamics \cite{2016ARA&A..54..491B,2020A&ARv..28....1L,2021NewAR..9301615R}. 
%Protostell jets can be observed in molecular, atomic ionized forms. Molecular jets, traced in multiwavelengths: radio, (sub)milimeter, infrared to optical wavelengths \cite{2016ARA&A..54..491B,2020A&ARv..28....1L,2021NewAR..9301615R}. by species such as CO, SiO, and SO, predominantly probe the denser, colder gas. (e.g., M. M. Dunham et al. 2014; L. Podio et al. 2021; L. S. Dutta et al. 2024).   Molecular hydrogen is one of the most common tracers of the jet in infrared wavelengths. Atomic and ionized jets, often seen in optical and near-infrared forbidden emission lines like [O I], [Fe II] [S II], trace hotter ((D. A. Neufeld et al. 1998; B. Lefloch et al. 2003; S. Maret et al. 2009; B. Nisini et al. 2010; T. Giannini et al. 2011, ., Ray 2021, T. P. Ray et al. 2023, van Dishoeck et al. 2025).  
The combination of molecular, atomic, and ionized jet studies provides a comprehensive view of jet evolution and propagation.  Observations of these multiwavelength species enable the estimation of crucial physical parameters. For instance, the jet mass-loss rate provides a measure of how efficiently the protostar expels material, while the jet and outflow momentum flux or force offers insights into the feedback exerted on the surrounding environment. The dynamical time inferred from jet length and velocity gives an estimate of the jet's age and episodic behavior.  Additionally, by comparing jet and outflow mass-loss rates to the estimated accretion rates, often inferred from bolometric luminosity or infall tracers, one can evaluate the efficiency of jet launching mechanisms and the coupling between mass ejection and accretion. Together, these parameters help to constrain the physical processes that govern jet launching, collimation, propagation, and their impact on star and disk evolution.

\begin{figure}%[H]  
    \includegraphics[width=0.8\linewidth]{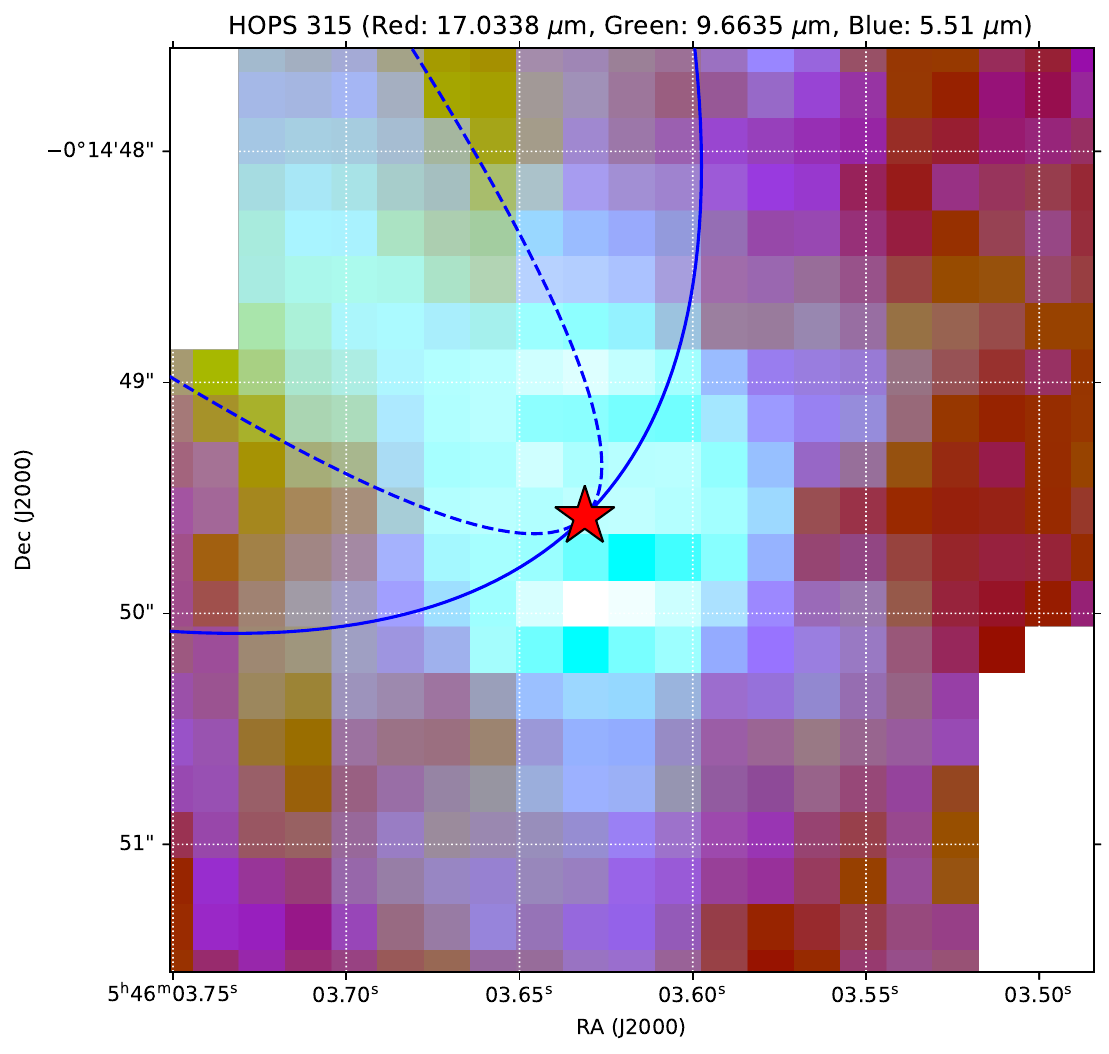}
\caption{JWST/MIRI color composite image of the jet/outflow system HOPS 315. The image has been produced using JWST GO Cycle 1 data (Proposal ID: 1854; PI: Melissa McClure), following methods described in \citet[][]{2025dutta}); red: H$_2$ 0-0 S(1) 17.0338~\textmu m; green: H$_2$ 0-0 S(3) 9.6635~\textmu m; blue: H$_2$ 0-0 S(7) 5.51~\textmu m. The outflow cavity is indicated by a solid parabola, the central axial jet by a dotted parabola, and the protostellar position is marked with an asterisk.}
    \label{fig:jwst_miri_hops315}
\end{figure}

Recent high-angular resolution observations with instruments such as the Atacama Large Millimeter/submillimeter Array (ALMA) and the James Webb Space Telescope (JWST) have revolutionized our understanding of protostellar jets and outflows. ALMA’s ability to resolve molecular emission at sub-arcsecond scales has revealed detailed structures of jet launching regions and outflow cavity walls (e.g., Figure \ref{fig:ALMA_jet_out_cont}), while JWST’s infrared sensitivity enables unprecedented views of shock-excited gas and embedded jet components (e.g., Figure \ref{fig:jwst_miri_hops315}). These advances provide new constraints on the jet launching mechanisms, evolutionary status of a protostar, accretion phase, and their impact on star and planet formation. In this review, we describe recent advances in the study of protostellar jets using observations from the ALMA and JWST telescopes.

%The introduction should briefly place the study in a broad context and highlight why it is important. It should define the purpose of the work and its significance. The current state of the research field should be reviewed carefully and key publications cited. Please highlight controversial and diverging hypotheses when necessary. Finally, briefly mention the main aim of the work and highlight the principal conclusions. As far as possible, please keep the introduction comprehensible to scientists outside your particular field of research. Citing a journal paper \citep{ref-journal}.  Now citing a book reference \citep{ref-book1,ref-book2} or other reference types \citep{ref-unpublish,ref-url}. Please use the command \citep{ref-proceeding,ref-thesis} for the following MDPI journals, which use author--date citation: Administrative Sciences, Arts, Behavioral Sciences, Businesses, Econometrics, Economies, Education Sciences, European Journal of Investigation in Health, Psychology and Education, Games, Genealogy, Histories, Humanities, Humans, IJFS, Journal of Intelligence, Journalism and Media, JRFM, Languages, Laws, Literature, Psychology International, Publications, Religions, Risks, Social Sciences, Tourism and Hospitality, Youth. 

%\section{From Early Observations to Ongoing Studies of Protostellar Jets}

\section{Discovery and Theoretical Foundations}
%\section{Advances in Protostellar Jet Studies}
%\subsection{Historical Milestones}

The discovery of protostellar jets began with the identification of compact emission-line nebulae near young stars, now known as Herbig--Haro (HH) objects. These were independently reported by Herbig (1951) and Haro (1952), who interpreted them as signatures of high-velocity outflows from protostars interacting with the surrounding interstellar medium \citep{1951ApJ...113..697H,1952ApJ...115..572H}. Their work laid the foundation for the modern study of stellar jets and outflows. Subsequently, \citet[][]{1980ApJ...239L..17S} detected large-scale bipolar molecular outflows traced by CO emission.  It is now well established that molecular outflows and jets are a ubiquitous feature of accreting, rotating, and magnetized protostellar systems \citep{
1980ApJ...239L..17S,1992A&A...261..274C,1996A&A...311..858B,2014ApJ...783...29D,2015A&A...576A.109Y,2021A&A...648A..45P,2024AJ....167...72D}.
These observations also revealed a strong link between outflow activity, accretion processes, and protostellar evolution. Together, these discoveries established that mass loss is a fundamental aspect of early stellar evolution.

%\subsection{Theoretical Framework and Jet Launching Mechanisms}

Protostellar jets and outflows emerge naturally during the gravitational collapse of dense molecular cloud cores. In the initial isothermal collapse phase, efficient cooling allows the core to contract while maintaining a temperature of typically $\sim$10~K. As central densities increase, the collapse becomes adiabatic, leading to the formation of the first hydrostatic core \citep{1969MNRAS.145..271L}. During this stage, magneto-centrifugal forces and magnetic pressure gradients launch low-velocity, wide-angle outflows \citep{1998ApJ...502L.163T,2008ApJ...676.1088M}. When temperatures reach $\sim$2000~K, molecular hydrogen dissociates, triggering a second collapse that forms the protostar. This leads to the emergence of fast, highly collimated jets driven by strong magnetic forces from the inner regions of the disk or the protostar itself \citep{2008ApJ...676.1088M,2013ApJ...763....6T}.

% \section{Observational Signatures of Disk Winds and Jets}
% Observations reveal that protostellar systems often exhibit both slow disk winds and fast, collimated jets. Disk winds originate from larger radii ($\sim 1$--10~au) in the disk and are typically launched via magneto-centrifugal processes. These winds show moderate velocities of $\sim 10$--50~km~s$^{-1}$ and broad opening angles \citep{blandford1982,Ferreira2006}. In contrast, highly collimated jets are launched from the inner regions of the disk ($<1$~au) or the immediate vicinity of the protostar. These jets reach much higher velocities, often exceeding 100~km~s$^{-1}$, and are tightly focused along the rotation axis \citep{frank2014, bally2016}. Observations in atomic lines (e.g., [O~\textsc{i}], [S~\textsc{ii}]) and molecular tracers (e.g., SiO, CO) have confirmed these distinct kinematic components. The coexistence of disk winds and jets indicates a layered outflow structure, where both components play essential roles in removing angular momentum and regulating accretion onto the protostar.

%%%%%%%%%%%%%%%%%%%%%%%%%%%%%%%%%%%%%%%%%%
\section{Physical Properties and Observational Characteristics of Jets \linebreak and Outflows}
%Section 2: Physical Properties and Observational Characteristics of Jets and Outflows
%
%This section can cover:
\subsection{Molecular vs Neutral and Ionized Components}
\begin{figure}%[H]  
    \includegraphics[width=0.72\linewidth]{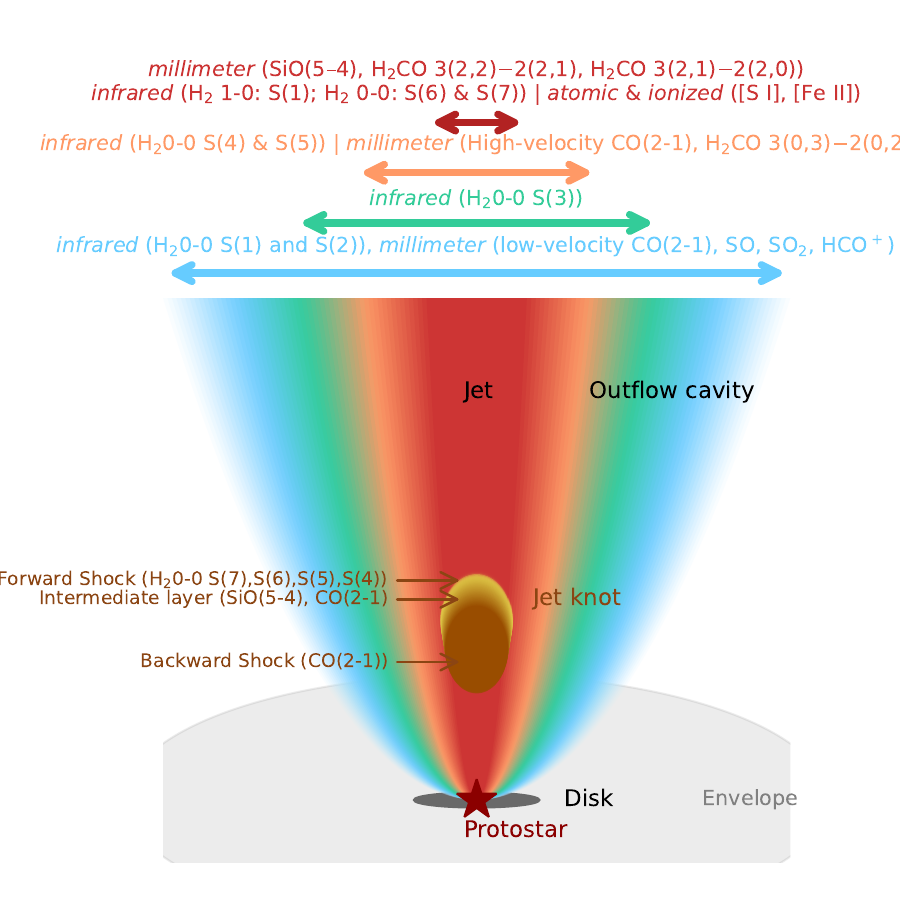}
    \caption{Schematic diagram showing the detection ranges of various molecular, atomic and ionic species in outflow--jet--shocks, reproduced and expanded from the data in \citet[][]{2025dutta} to include additional species. The distribution of individual species are indicated along the top axis by bidirectional arrows, while the corresponding colors illustrate their relative spatial extents as depicted in \mbox{the schematic.}
    }
    \label{fig:schematic_molecular_atomic}
\end{figure}

Jets from YSOs consist of molecular, neutral atomic, and ionized components, each traced by distinct sets of emission lines that probe different physical conditions and spatial regions within the jet–outflow system, as demonstrated in Figure \ref{fig:schematic_molecular_atomic}. In the case of (sub)millimeter observations, molecular tracers such as CO, SiO, SO, and H$_2$CO are effective in probing the cooler components of the flow, typically at temperatures ranging from a few tens to a few hundreds of Kelvin. The lower-$J$ transitions, such as CO(1–0), CO(2–1), SiO(2–1), SiO(3–2), and H$_2$CO lines, are sensitive to lower-density, low-velocity, wide-angle outflows, or wind components, and are often associated with entrained ambient gas (e.g., \citep[]{1996A&A...311..858B,2014ApJ...783...29D}). However, some of these transitions can also be excited in high-velocity, high-density jets, particularly in shock-excited regions. In contrast, the higher-excitation transitions, including CO(3–2), SiO(5–4), SiO(8–7), and high-$J$ H$_2$CO lines, trace denser, warmer gas in the collimated jet cores, where shocks and magnetic launching mechanisms dominate the dynamics (e.g., \citep[]{2014ApJ...783...29D,2015A&A...576A.109Y,2017NatAs...1E.152L,2021A&A...648A..45P,2024AJ....167...72D}). In infrared observations, the H$_2$ pure rotational (e.g., H$_2$ 0-0 S(1)) and ro-vibrational  (H$_2$ 1-0 S(1), S(2), S(3), S(4), S(5), S(7, S(7)) are frequently observed in jets and outflows (e.g., Figure \ref{fig:jwst_miri_hops315}). The shorter wavelengths (e.g., H$_2$: 0-0 S(1); 1-0 S(7), S(6), S(5), S(4)) are often concentrated in the jets, while the longer wavelengths are also trace the wide angle outflow components (e.g., 1-0 S(1), S(2), S(3)) \cite{2023ApJ...951L..32H,2023Natur.622...48R,2024ApJ...962L..16N,2024A&A...687A..36T,2024A&A...688A..26A,2024A&A...691A.134C,2025A&A...695A.145V,2025ApJ...982..149O,2025ApJ...985..225L}. % 2025arXiv250508002V 
 We note that, in some cases, the shorter-wavelength, lower-excitation rotational lines of H$_2$ also trace the wider-angle outflow rather than the collimated jet, as demonstrated by the JOYS program \citep{2025A&A...699A.361V}; however, this behavior is not universal.

In contrast, neutral atomic (e.g., [O I], [C I], [S I]) and ionized atomic tracers (e.g., [S II], [N II], [O III], [Fe II], H$\alpha$) reveal hotter ($T \sim 10^4$~K), lower-density regions, often linked to fast-moving, collimated jets and internal shocks \cite{1995ApJ...452..736H,2016ARA&A..54..491B}.  Forbidden lines such as  [S\,\textsc{ii}]~$\lambda$6716, 6731~\AA, [N\,\textsc{ii}]~$\lambda$6583~\AA, and [Fe\,\textsc{ii}]~1.644\,\textmu m commonly serve as effective tracers of physical conditions such as electron density, ionization fraction, and excitation temperature \citep{2001ARA&A..39..403R,2006A&A...459..821G}.  %The [S\,\textsc{ii}] doublet, for example, is widely used to estimate electron densities in the range $10^2$–$10^4$\,cm$^{-3}$ based on its line ratio \citep{1989agna.book.....O}. 
Hydrogen recombination lines like H$\alpha$ are also detected in ionized jets and accretion regions, indicating regions of high temperature and density \citep{1998ApJ...492..743M}.
Moreover, mid-infrared ionized lines such as [Ne\,\textsc{ii}]~12.8\,\textmu m and [Fe\,\textsc{ii}]~25.99\,\textmu m, observed with instruments like \textit{Spitzer} and \textit{JWST}, provide access to deeply embedded or heavily extincted jet regions \citep{2007ApJ...665..492L,2012A&A...545A..44P}. %Lahuis2007, Podio2012, Ustamujic2023}.

Spatially, molecular emission tends to dominate on larger scales where shocks dissipate, while ionized emission is prominent near jet-driving sources or in high-velocity knots along the jet axis. Together, these components offer complementary insights: molecular lines trace momentum transfer and jet-envelope interactions, while ionized lines reveal shock heating, excitation mechanisms, and jet launching dynamics. A combined analysis is essential for a complete understanding of jet physics across evolutionary stages.

\vspace{-12pt}

\subsection{Morphology and Structure}
\begin{figure}%[H]  
    \includegraphics[width=1\linewidth]{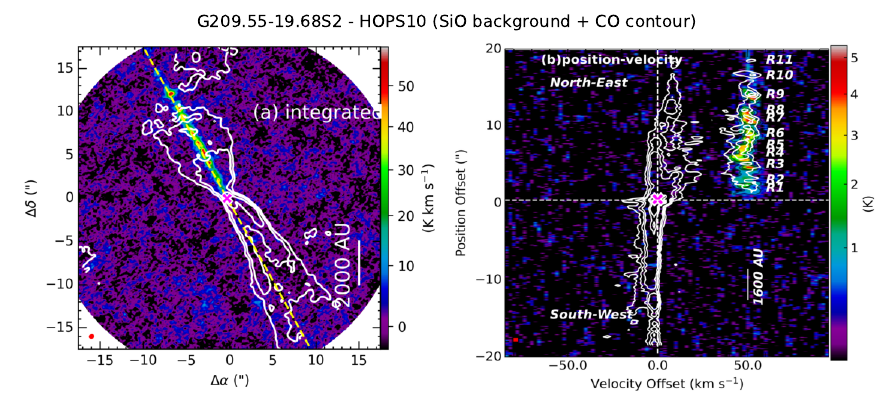}
\caption{
ALMA SiO and CO maps of the protostar G208.55$-$19.68S2 (HOPS 10), showing episodic knots and the monopolar nature of the jets at $\sim$150 AU resolution, adapted from \mbox{\citet[][]{2024AJ....167...72D}.} 
(\textbf{a}) integrated SiO emission map (background), with CO emission contours overlaid in white. 
\mbox{(\textbf{b}) Position–velocity} diagram with SiO emission as the background and CO contours in white. The location of the knots are marked with R1, R2, \ldots, R11.  The mean deprojected jet velocity is estimated to be $\sim 146^{+47}_{-46}$ km~s$^{-1}$, assuming an inclination angle of $\sim 20^{+10}_{-5}$ degree \citep[][]{2024AJ....167...72D}.
}
\label{fig:monopolarity_episodicJet}
\end{figure}

%    Describe the observed shapes, collimation, jet knots, bow shocks, outflow cavities, and how morphology varies with source age or mass.
%\subsection{Morphology and Structure: Jets vs Outflows}
Jets and outflows from protostars exhibit distinct morphological and structural characteristics, reflecting their different physical origins at the disk, their interaction with the envelope, and the evolutionary stages of the driving protostars. 
Collimated jets are typically narrow with high velocities, $v_j \sim 40$ km\,s$^{-1}$ to a few 100 s of km\,s$^{-1}$ with a mean of $\sim$110 km\,s$^{-1}$ \citep{2024AJ....167...72D}. As shown in the example in Figure \ref{fig:monopolarity_episodicJet}, the jets often appear as chains of bright knots or bow shocks along their axes (e.g., \cite[]{2001ARA&A..39..403R,2015Natur.527...70P,2016ARA&A..54..491B,2024AJ....167...72D}). These knots result from episodic ejection events and internal shocks, while bow shocks mark the interaction between the jet and the surrounding medium. Jets are usually well-collimated with opening angles of only a few degrees, particularly near the launching region. In contrast, molecular outflows are broader, wide-angle structures with lower characteristic velocities \mbox{($v \lesssim$ 10--30 km\,s$^{-1}$)} and exhibit a more conical or lobe-like appearance \cite{2007prpl.conf..245A,2014prpl.conf..451F,2024AJ....167...72D}.

The intrinsic opening angle ($\theta_{\text{int}}$) of the wide-angle outflow can be obtained from the observed opening angle, $\theta_{\text{obs}}$, and the inclination angle, $i$ (with $i = 0^\circ$ representing an edge-on disk and $i = 90^\circ$ a pole-on system), through the geometric relation:  
\begin{equation}
\theta_{\text{int}} = 2 \cdot \tan^{-1} \left( \frac{\tan\left( \tfrac{\theta_{\text{obs}}}{2} \right)}{\cos(i)} \right).
\end{equation} 
Protostellar outflows trace ambient material that has been entrained and accelerated by underlying jets or wide-angle winds, often producing cavities shaped by shocks. 
%Their morphology frequently reflects the evolutionary state of the source (e.g. \citep[]{1996A&A...311..858B,2014ApJ...783....6V,2017AJ....153..173H}.
%: Class~0 systems typically drive massive, poorly collimated flows, while Class~I and II objects exhibit narrower cavities and well-collimated jets .  
%Outflow cavities, carved into the envelope by bow shocks or wide-angle winds, are prominent manifestations of this process \cite{2007prpl.conf..245A,2014ApJ...783....6V}. They are commonly observed in scattered infrared light and in molecular line tracers such as CO, which delineates the cavity walls. 
The cavity opening angle serves as a useful indicator of protostellar evolution: in the Class~0 phase, cavities are narrow and poorly defined due to dense envelopes and high accretion \mbox{rates \cite{1996A&A...311..858B,2014ApJ...783....6V,2017AJ....153..173H};} during the Class~I stage, successive ejection events progressively widen the cavities as the envelope mass declines; by Class~II, cavities often appear broad and conical, with some systems revealing the underlying disk--jet structure \cite{2008ApJ...675..427S,2011ApJ...743...91O}.

%Morphology also varies with protostellar mass. Low-mass protostars tend to drive well-collimated jets and molecular outflows, whereas intermediate- and high-mass protostars often show less collimated, more chaotic outflow structures due to stronger radiation pressure, ionization, and clustering effects \cite[e.g.,][]{2002ApJ...576..222B}. However, high-resolution observations increasingly reveal collimated jets even in some high-mass systems, suggesting that accretion-driven jet mechanisms may be universal, though modified by environmental conditions \cite{2011ApJ...740..107C}. Thus, the morphology of jets and outflows encodes vital information about both the evolutionary stage and the mass of the driving source, making it a key tool for studying star formation across the mass spectrum.

%\subsection{Kinematics and Dynamics}
%    Present typical velocities, velocity gradients, rotation signatures, acceleration/deceleration, and jet wiggling/precession.
\subsection{Shock Processing}

Shocks in protostellar jets are commonly classified as either J-type (jump) or C-type (continuous), which differ in their internal structure and in how they affect molecules and chemistry  (e.g., \citep[]{1997IAUS..182..181H}).  In a J-type shock, the gas undergoes an abrupt discontinuity in physical conditions  (density, temperature) at the shock front; molecules may be dissociated or ionized in  the shock, and they must re-form in the post-shock region. In contrast, a C-type shock  develops in a weakly ionized medium with a significant magnetic field: ions (and charged  particles) are tied to magnetic field lines and drag neutrals via ion--neutral collisions,  resulting in a smoother, continuous transition in physical conditions (rather than a  sharp ``jump'') over a finite shock thickness. Because of this, molecules may survive  passage through a C-type shock, and heating is more gradual; these differences strongly  influence emission signatures, chemical abundances, and cooling behavior 
(e.g., \citep[]{1997IAUS..182..181H}).  In many jet/outflow systems, a mixture of J-type and C-type shock zones may coexist  (e.g., at bow shock apices versus wings) depending on local physical conditions.

Protostellar jets are inherently variable, and fluctuations in their ejection velocity naturally generate internal shock structures where faster parcels of gas collide with slower ones. As demonstrated in Figure \ref{fig:schematic_molecular_atomic}, these interactions form compact internal working surfaces characterized by a reverse shock that decelerates the jet gas and a forward shock that accelerates the ambient medium \citep{1990ApJ...364..601R}. Numerical simulations and observations \mbox{(e.g., \citep[]{2017A&A...597A.119T})} demonstrate that such internal shocks not only act along the jet axis but also drive sideways ejections, pushing shocked gas laterally into the surrounding envelope and producing expanding bow-shock structures within the jet. On larger scales, the overall bow-shaped interaction between the jet and its environment is described by a system of forward and reverse shocks \citep{2001ApJ...557..429L,2021ApJ...909...11J}: the forward shock propagates into the ambient cloud, sweeping up molecular gas into wide cavities, while the reverse shock propagates back into the jet beam, compressing and energizing the jet material. This combination of internal shocks with lateral expansion and terminal forward--reverse shock pairs provides a unifying framework for the stratified kinematic and chemical structures observed in protostellar outflows, including compact knots and bullets along the jet channel together with broader, slower outflow lobes and shells.

\subsection{Chemical Composition in Shock Signatures}

Jets and outflows profoundly alter the chemistry of their surroundings through shock-driven processes. As high-velocity material impacts the ambient medium, shocks trigger heating, sputtering of dust grains, and gas-phase reactions that release molecules from grain mantles, thereby enhancing the abundances of key species \cite{1983ApJ...264..485D,1997ApJ...487L..93B}. Prominent tracers of such activity include submillimeter molecular species such as SiO, SO, CO, and CH$_3$OH whose abundances can increase by several orders of magnitude compared to quiescent \mbox{gas \cite{2008A&A...490..695G,2010A&A...522A..91T},} as well as infrared atomic and ionic lines (e.g., [Fe\,II], [S\,I]) and molecular transitions at various vibrational or rovibrational levels.

SiO is widely regarded as a hallmark of strong shocks (shock velocity, $v_s \gtrsim 20$ km\,s$^{-1}$), originating from the sputtering of silicate grains followed by gas-phase reactions. In contrast, molecules such as CH$_3$OH and H$_2$CO are typically associated with lower-velocity shocks or thermal desorption of grain mantles \cite{1997A&A...321..293S,1991ApJ...373..254G}. As illustrated in Figure~\ref{fig:schematic_molecular_atomic}, different tracers highlight distinct layers of the shock structure: CO transitions trace the extended shock region, higher-$J$ SiO transitions are strongest at the forward shock and its immediate post-shock layers, while near-infrared H$_2$ emission marks the shock front.  

% These chemical signatures also act as chemical clocks, providing constraints on shock timescales and outflow evolution. Certain molecules reach peak abundances shortly after the shock passes and subsequently decline as they are destroyed or re-adsorbed onto grains \cite{Jimenez2005}. Spatial variations in molecular stratification within jets and outflows thus offer a record of shock ages and propagation histories.  

These chemical signatures also act as chemical clocks, constraining shock timescales and outflow evolution. For example, models predict that elevated H$_2$O abundances induced by C‐type shocks persist for $4\text{–}7\times10^5\,$yr before declining again \cite{1998ApJ...499..777B}. MHD shock models and gas–grain chemistry suggest that molecules sputtered from grain mantles may survive for only tens of years post‐shock, showing rapid decline in abundance \cite{2018Ap&SS.363..151N}. Observations of complex species in the L1157-B1 outflow, whose age is $<2000\,$yr, indicate that those molecules are shock-released mantle products whose presence traces recent shock \mbox{activity \cite{2008ApJ...681L..21A}.} Models of episodic protostellar accretion likewise show that molecular tracers such as HCO$^+$ and N$_2$H$^+$ return to quiescent levels only after $10^3\text{–}10^5\,$yr, making them effective clocks for past heating or shock events \cite{2015A&A...577A.102V}.

On larger scales, repeated shocks inject turbulence, shape photodissociation regions (PDRs), and modify the ionization structure of the surrounding cloud \cite{2007prpl.conf..245A,2021A&A...648A..24V}. Over time, such chemical feedback enriches outflow cavities and influences the conditions for subsequent star formation. Studying the chemical composition and shock tracers in protostellar jets therefore provides critical insights into the coupling between stellar feedback, gas dynamics, and astrochemistry in star-forming environments.

%%%%%%%%%
\subsection{Velocity Gradients and Rotation Signatures in Protostellar Jets and Outflows}

% Velocity gradients perpendicular to the jet or outflow axis are commonly observed in high-resolution observations of protostellar systems. These gradients have been interpreted as possible signatures of jet rotation \citep{2002ApJ...576..222B, 2004ApJ...604..758C,2007ApJ...670.1188L, 2017NatAs...1E.152L}. The detection of such gradients offers insight into the angular momentum transport mechanisms at the earliest stages of star formation.

% In several Class 0/I protostellar jets, transverse velocity shifts on the order of a few km~s$^{-1}$ across the jet width have been reported, suggesting rotation in the launching region. For example, \citet{2002ApJ...576..222B} observed systematic velocity gradients across the DG Tau jet using HST/STIS, interpreting them as rotational motion. ALMA observations have extended such detections into molecular jets, such as SiO and CO, in deeply embedded protostars \citep{2007ApJ...665..492L,2017NatAs...1E.146H}. \citet{2017NatAs...1E.152L} have estimated the jet launching radius in HH 212 system using very high- angular resolution ($\sim$ 16 AU) observations of jet, which was $\sim  0.05^{+0.05}_{-0.02}$ AU, which was best possible estimate so far. 

Velocity gradients perpendicular to the jet axis are commonly observed in high spectral and spatial resolution studies of protostellar systems. These gradients have been interpreted as potential signatures of jet rotation \citep{2002ApJ...576..222B, 2004ApJ...604..758C, 2007ApJ...670.1188L, 2017NatAs...1E.152L}, providing insights into angular momentum transport during the earliest stages of star formation. If interpreted as rotation, these gradients allow estimates of the jet launching radius under the magneto-centrifugal wind framework (e.g., \citep[]{2003ApJ...590L.107A,2006A&A...453..785F,2017NatAs...1E.152L}).
%
%, using the relation (e.g. \citep[]{2003ApJ...590L.107A,2006A&A...453..785F,2017NatAs...1E.152L}
%:
%\[
%r_0 \approx \frac{r_{\perp} v_{\phi}}{v_p}
%\]
%where \( r_0 \) is the launching radius, \( r_{\perp} \) is the observed distance from the jet axis, \( v_{\phi} \) is the rotational velocity, and \( v_p \) is the poloidal (outflow) velocity. Measured rotational signatures suggest launching radii of a few au, consistent with disk-wind models. 
However, confirmation of rotation requires careful consideration of projection effects, beam smearing, and asymmetric shock structures, as emphasized by \citet{2005A&A...435..125S}.

In several Class 0/I protostellar jets, transverse velocity shifts of a few km~s$^{-1}$ across the jet width have been reported, suggesting rotation in the launching region. For instance, \citet{2002ApJ...576..222B} observed systematic velocity gradients across the DG Tau jet using HST/STIS, interpreting them as rotational motion. ALMA observations have extended such detections to molecular jets, including SiO and CO, in deeply embedded \mbox{protostars \citep{2007ApJ...665..492L, 2017NatAs...1E.146H}.} In the HH 212 system, \citet{2017NatAs...1E.152L} estimated the jet launching radius to be $\sim 0.05^{+0.05}_{-0.02}$~AU using very high angular resolution ($\sim$16~AU) observations, providing the most precise measurement to date.

In wide-angle molecular outflows, transverse velocity gradients have also been identified, although distinguishing rotation from other asymmetries (e.g., jet precession, sideways ejection) remains challenging \citep{2007prpl.conf..245A,2009A&A...494..147L}. Detailed modeling is often required to rule out alternative explanations.  Rotating protostellar outflows trace angular momentum removal from disks, often via magnetohydrodynamic (MHD) disk winds. For example, in HH~212, SO emission reveals a rotating wide-angle disk wind around an episodic SiO jet. Resolving this outflow shows a collimated jet, a wide-angle disk wind, and a jet-driven cavity, confirming that disk winds efficiently extract angular momentum and providing a probe of large-scale magnetic fields in disks \citep{2021ApJ...907L..41L}. Recent ALMA  observations of HH~212 at $\sim$24~au \mbox{resolution \citep{2024ApJ...977..126L}} reveal a rotating wind launched from 9--15~au, situated between the SiO and CO shells. This shocked rotating wind interacts with the inner X-wind and mixes with SO emission, efficiently removing angular momentum and shaping the outflow structure.

\subsection{Physical Parameters}
%    Mass-loss rates, momentum flux, energetics, magnetic field measurements, densities, temperatures.

Jets and outflows from protostars exhibit a wide range of physical parameters that are crucial for understanding their roles in star formation. Typical mass-loss rates range from $\sim$$10^{-9}$ to $10^{-5}$\,M$_\odot$\,yr$^{-1}$, and generally vary with protostellar mass and evolutionary stage \cite{2007A&A...468L..29C,2014prpl.conf..451F}.
Although there is a connection between mass-loss rates and protostellar mass and evolution, this relationship is not monotonic; for instance, individual systems can vary significantly and do not always follow a trend of higher mass-loss rates with increasing protostellar mass.  Molecular outflows usually dominate in terms of mass, while atomic jets carry significant kinetic energy and momentum flux. The momentum flux (or force) of outflows can reach values of $10^{-6}$ to $10^{-2}$\,M$_\odot$\,km\,s$^{-1}$\,yr$^{-1}$, often correlating with the luminosity and accretion rate of the driving source \cite{2002ApJ...576..222B, 2007prpl.conf..245A}. Energetics analyses reveal that outflows can inject enough energy and momentum to significantly affect their surroundings, regulating star formation efficiency and driving turbulence in \mbox{molecular clouds.}

Typical particle densities vary from $10^3$ to $10^6$\,cm$^{-3}$ in molecular outflows to \mbox{$10^2$ to $10^4$\,cm$^{-3}$} in atomic jets, with local enhancements in shock regions \cite{1996A&A...311..858B,1995ApJ...452..736H}. Corresponding temperatures range from $\sim$10 to 100\,K in cold molecular gas up to several thousand Kelvin in ionized or shocked atomic components \cite{2005A&A...441..159N, 2013ApJ...778...71G}. Together, these physical parameters provide a comprehensive framework for evaluating jet energetics, mass and momentum budgets, and their feedback on star-forming environments.

\begin{figure}%[H]  
    \includegraphics[width=0.8\linewidth]{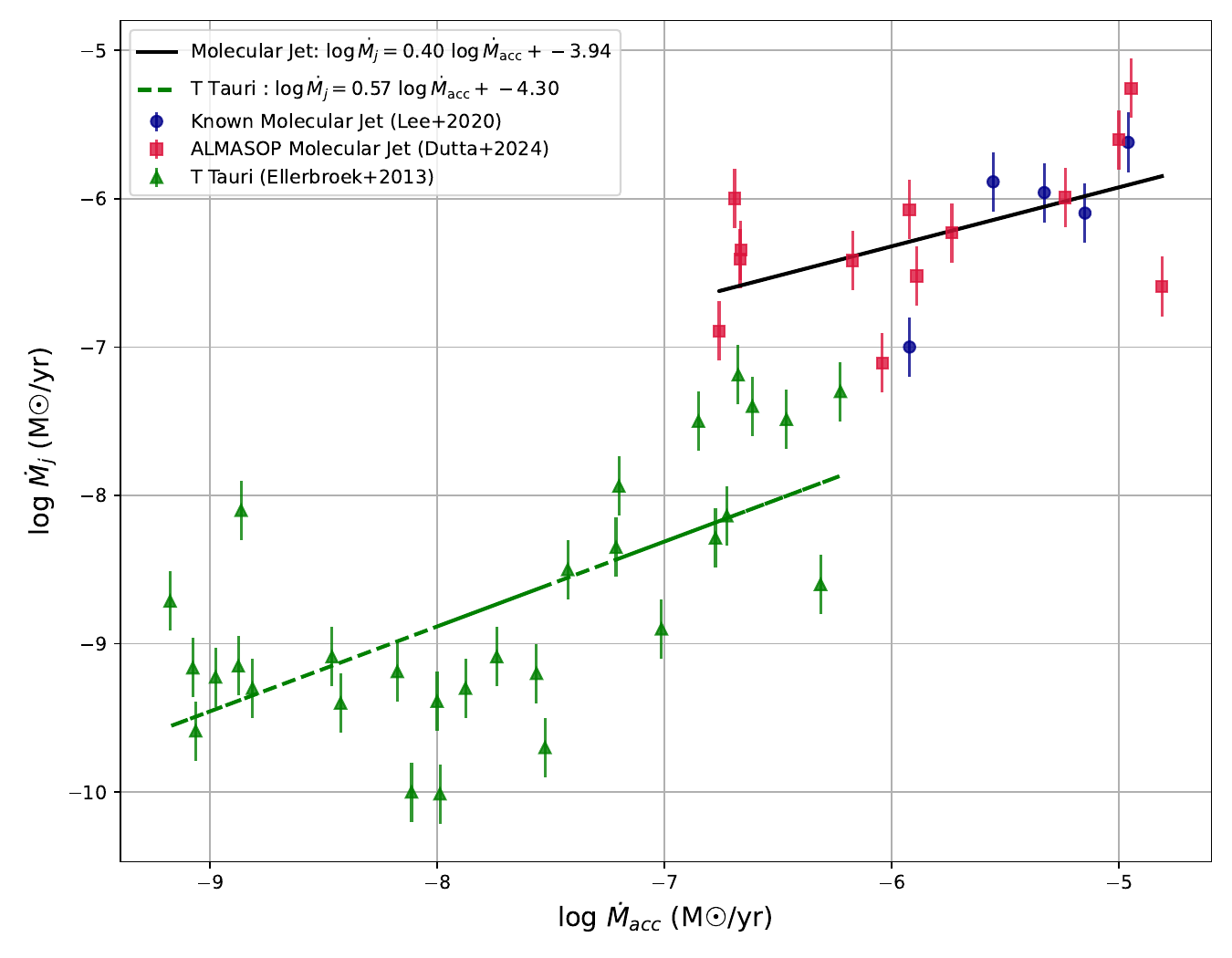}
    \caption{Observed jet mass-loss rate ($\dot{M}j$) versus accretion rate ($\dot{M}{\rm acc}$) for protostellar jet sources. Blue symbols represent well-studied protostars with molecular jets from \citet{2020A&ARv..28....1L}, while red symbols denote molecular jets in the ALMASOP sample from \citet{2024AJ....167...72D}. Green symbols show  more evolved  T Tauri sources from \citet{2013A&A...551A...5E}. Separate fits are shown for molecular jets (black line) and T~Tauri stars (green line). All the data in this plot were adopted with permission of the lead authors.
    }
    \label{fig:jet_accretion}
\end{figure}

Figure~\ref{fig:jet_accretion} shows the jet mass-loss rates as a function of accretion rates for protostars. Sources with molecular jets, which are typically younger, exhibit systematically higher accretion and ejection activity than T~Tauri stars, which are generally more evolved and show lower rates. The two groups have been fitted separately with a logarithmic relation:  

\begin{equation}
\log_{10}(\dot{M}_j) = \alpha \cdot \log_{10}(\dot{M}_{\text{acc}}) + \beta ,
\end{equation}
\begin{equation}
\dot{M}_j \propto \dot{M}_{\text{acc}}^{\alpha}
\end{equation}

The molecular jets show relatively shallow slopes ($\alpha$$\sim$0.40) compared to T~Tauri stars ($\alpha$$\sim$0.57). 
This implies that younger objects tend to accrete at higher rates but eject smaller fraction of that mass. This Figure ~\ref{fig:jet_accretion}  illustrates that jets and outflows play a significant role in regulating protostellar growth and evolution. In particular, as the system evolves toward the T Tauri stage, the relative mass-loss efficiency increases, indicating that outflow activity becomes more effective at removing material rather than settling the mass onto \mbox{the star.}
 
\subsection{Monopolarity Nature}

While many protostellar jets and outflows exhibit bipolar symmetry, an increasing number of systems show monopolar or highly asymmetric morphologies. In such cases, emission is detected from only one lobe---in molecular lines, atomic tracers, or continuum maps. For example, the G209.55$-$19.68S2 source (HOPS~10) in Figure~\ref{fig:monopolarity_episodicJet} shows one-sided SiO jet emission, with the opposite lobe absent in both SiO and CO.  

The observed jet asymmetry likely arises from multiple factors, with geometry being key. In protostars, JWST-detected infrared emission is strongly affected by extinction from the envelope (e.g., Figure \ref{fig:jwst_miri_hops315}), particularly along the redshifted side (e.g., \citep[]{2023ApJ...951L..32H,2024A&A...692A.143B,2024ApJ...966...41F,2025A&A...699A.361V}). The orientation relative to the line of sight also introduces projection effects, while deeply embedded regions may remain invisible due to high optical depth.  

Doppler boosting, important in relativistic AGN jets (e.g., M87; \citep[]{2007ApJ...668L..27K,2007ApJ...658..232C}), is negligible in protostellar jets since their velocities (a few hundred km\,s$^{-1}$) are too low. Yet nearly half of protostars with SiO jets appear monopolar at millimeter wavelengths \citep[]{2024AJ....167...72D}.
 Because dust extinction is minimal at submillimeter wavelengths, this high incidence cannot be explained by extinction alone, suggesting intrinsic launching asymmetries, possibly linked to magnetic field geometry or star--disk interactions.  

Jet-launching models that include a stellar magnetosphere generate oppositely directed poloidal fields and a quadrupolar toroidal structure, enabling reconnection and driving jets in both hemispheres---consistent with bipolar outflows (e.g., \citep[]{1994ApJ...429..781S,2000prpl.conf..789S,2003ApJ...599..363A,2007prpl.conf..277P}). In contrast, disk-only magnetic fields are inherently asymmetric: the toroidal field remains unipolar within each hemisphere, lacking reversals across latitudes. Reconnection and amplification processes, such as avalanche accretion streams, then operate effectively only in one hemisphere \citep[][]{2025ApJ...988..107T,2025arXiv250611333T}. This asymmetry favors jet launching on a single side, producing unipolar rather than bipolar outflows.  

%%%%%%%%%%%%%%%%%%%%%%%%%%%

\subsection{Episodic Nature and Their Origin}

Jets and outflows from protostars often display a distinctly episodic character, with chains of knots, bow shocks, and discrete ejections tracing their axes \citep{2001ARA&A..39..403R,2016ARA&A..54..491B,2015Natur.527...70P}. An example of such episodic knots is shown in Figure~\ref{fig:monopolarity_episodicJet}. These structures arise when variable ejection speeds cause faster material to overtake earlier, slower ejecta, generating internal \mbox{shocks \citep{1990ApJ...364..601R,1995ApJ...452..736H}.} Observations of proper motions and radial velocities suggest recurrence timescales ranging from a few years to several hundred years, depending on the \mbox{system \citep{2014A&A...563A..87E,2015ApJ...805..186L,2024AJ....167...72D}.}

This episodicity is widely attributed to unsteady accretion, with mass inflow onto the protostar closely linked to jet launching (e.g., \citep[]{2014prpl.conf..387A}). Proposed drivers include disk instabilities—gravitational or magneto-rotational—magnetospheric reconnection, and dynamical perturbations from companions \citep{2015ApJ...805..115V}. Disk or jet precession may further modulate the ejection geometry, producing complex morphologies \citep{2002ApJ...580..950M,2010ApJ...713..731L}. Such discontinuities likely reflect temporal variations in jet velocity or density, tied to quasi-periodic perturbations of the accretion flow. Candidate mechanisms include binary-induced variability, gravitational instabilities in the envelope or disk, episodic planetesimal accretion, and instabilities near dust sublimation fronts \citep{2014prpl.conf..387A,2020A&ARv..28....1L,2023ASPC..534..355F}.

\begin{figure}%[H]  
    \includegraphics[width=1\linewidth]{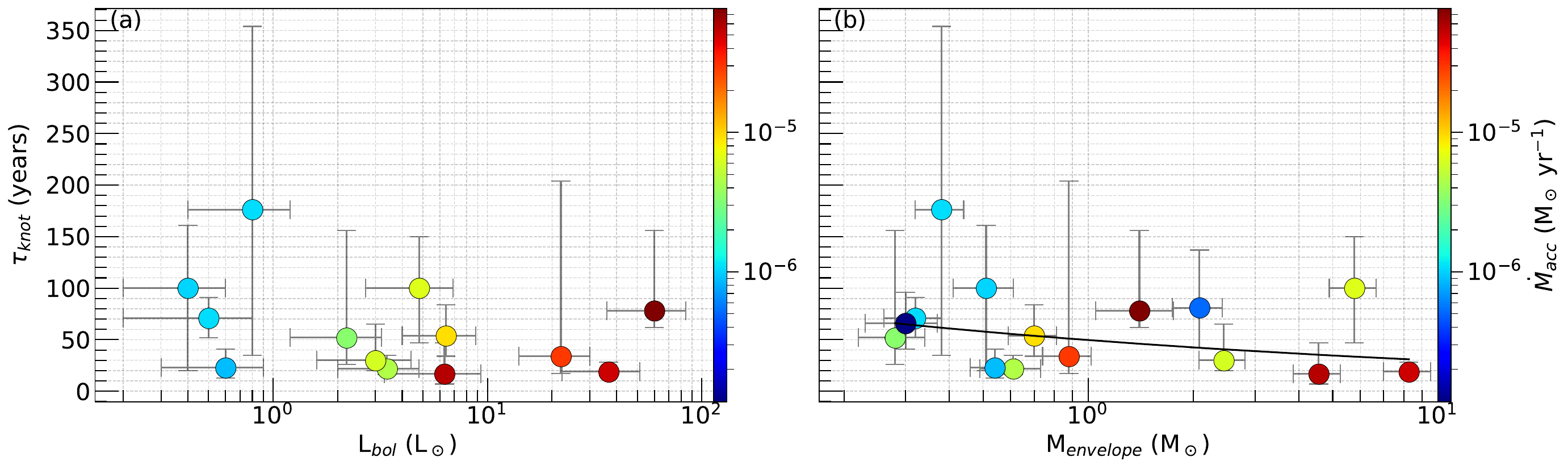}
    \caption{Average knot ejection intervals ($\tau_{knot}$) are displayed as a function of (\textbf{a}) bolometric luminosity (L${bol}$) and (\textbf{b}) envelope mass (M$_{envelope}$), reproduced using data from \citet[][]{2024AJ....167...72D} for consistency with this review. The color scale corresponds to the mass accretion rate ($\dot{M}_{acc}$). In panel (\textbf{b}), the straight line shows the best linear fit to the data.
    }   
    \label{fig:episodicity_luminosity_envelopeMass}
\end{figure}

Observationally, SiO jets often exhibit clumpy, knotty structures that may reflect quasi-periodic ejections, with inferred recurrence times of $\sim$20--175 years \citep{2024AJ....167...72D}. As shown in Figure \ref{fig:episodicity_luminosity_envelopeMass}, the episodicities do not show any clear dependence on luminosity, most likely because luminosity alone does not reliably trace evolutionary status. Protostellar luminosity is the sum of internal luminosity and variable accretion-driven contributions, the latter decreasing with time after an outburst. In contrast, envelope mass more directly reflects the evolutionary state of the core hosting the protostar (e.g., \citep[]{2023ApJ...944...49F}). 
%Systems with smaller envelope masses are therefore more likely to exhibit reduced accretion/ejection activity. Indeed, a mild anticorrelation is found between the ejection knot variability timescale and envelope mass. However, cores with inherently low-mass envelopes could not be distinguished or accounted for in this scenario.
Systems with smaller envelope masses tend to exhibit reduced accretion and ejection activity.  However, the relationship between ejection variability and envelope mass is not straightforward: the timescales traced by shocked knots indicate that accretion is highly system-dependent, influenced by factors such as the disk and stellar magnetic fields, as well as the infall properties of each individual system.  Cores with inherently low-mass envelopes could not be distinguished or fully accounted for in this scenario.

Multiple periodicities have been reported in individual systems, such as HH\,34, HH\,111, and HH\,212, spanning years to decades  \citep{1998Natur.394..862Z,2002A&A...395..647R}, suggesting the coexistence of different accretion perturbation modes. Complementary evidence comes from long-term monitoring: the JCMT Transient Survey revealed submillimeter variability on decadal scales \citep{2017ApJ...849...43H,2021ApJ...920..119L}, while NEOWISE mid-infrared data uncovered secular variability among embedded YSOs \citep{2021ApJ...920..132P}.  

Estimating orbital radii from observed knot periods yields characteristic perturbation zones at $\sim$2--25 au for typical protostellar masses \citep{2024AJ....167...72D}, consistent with instabilities in small to intermediate disks rather than binary orbital forcing. Although correlations with luminosity and envelope mass remain weak or statistically inconclusive, these episodic jets provide crucial probes of time-dependent accretion and the dynamical evolution of protostellar disks. Continued monitoring and modeling will be essential to clarify the link between accretion variability and jet launching.

\subsection{Disk Winds Around the Jets?}
Protostellar outflows generally consist of two main components: disk winds and jet winds (or simply jets), both of which play crucial roles in the star formation process. Disk winds are launched from a relatively extended region of the protostellar disk, typically ranging from $\sim$0.1 to a few astronomical units (au), where large-scale magnetic fields enable the ejection of material via magneto-centrifugal processes \citep{1982MNRAS.199..883B,2007prpl.conf..277P}. These winds tend to have broad opening angles and moderate velocities, often around 1 to a few \mbox{10 s~km~s$^{-1}$,} and are thought to efficiently carry away angular momentum from the disk, allowing sustained accretion. In contrast, jet winds originate from the innermost regions near the star--disk interface, typically within a fraction of an au. These jets are highly collimated along the flow-axis, with velocities that can exceed 40--300~km~s$^{-1}$, and are commonly modeled as magnetically driven X-winds or magnetospheric ejections \citep{1994ApJ...429..781S,2006A&A...453..785F}. Observationally, disk winds are often traced by molecular emission lines such as CO and SO at (sub)millimeter wavelengths, and by infrared lines of H$_2$ or atomic species in warmer regions (e.g., \citep[]{2016Natur.540..406B,2025dutta}). Jets, on the other hand, are traced by high-velocity atomic and ionized lines such as [S~II], [Fe~II], and H$\alpha$ in optical and infrared observations, and by shock-excited molecules such as SiO and CO in deeply embedded sources \citep{2014prpl.conf..451F,2016ARA&A..54..491B}. While both components are magnetically driven, they differ in terms of launching regions, kinematics, collimation, and chemical tracers. Together, they reveal a complex, multi-component ejection process, with jets often showing episodic variability and bow shocks, while disk winds drive more continuous, wide-angle flows. Recent high-resolution observations with ALMA and JWST are beginning to resolve these components simultaneously, offering new insights into their interplay and their roles in disk evolution and star formation feedback \citep{2017NatAs...1E.152L,2020A&A...640A..82T}.

\section{Launching Models}

\begin{figure}%[H]  
    \includegraphics[width=1\linewidth]{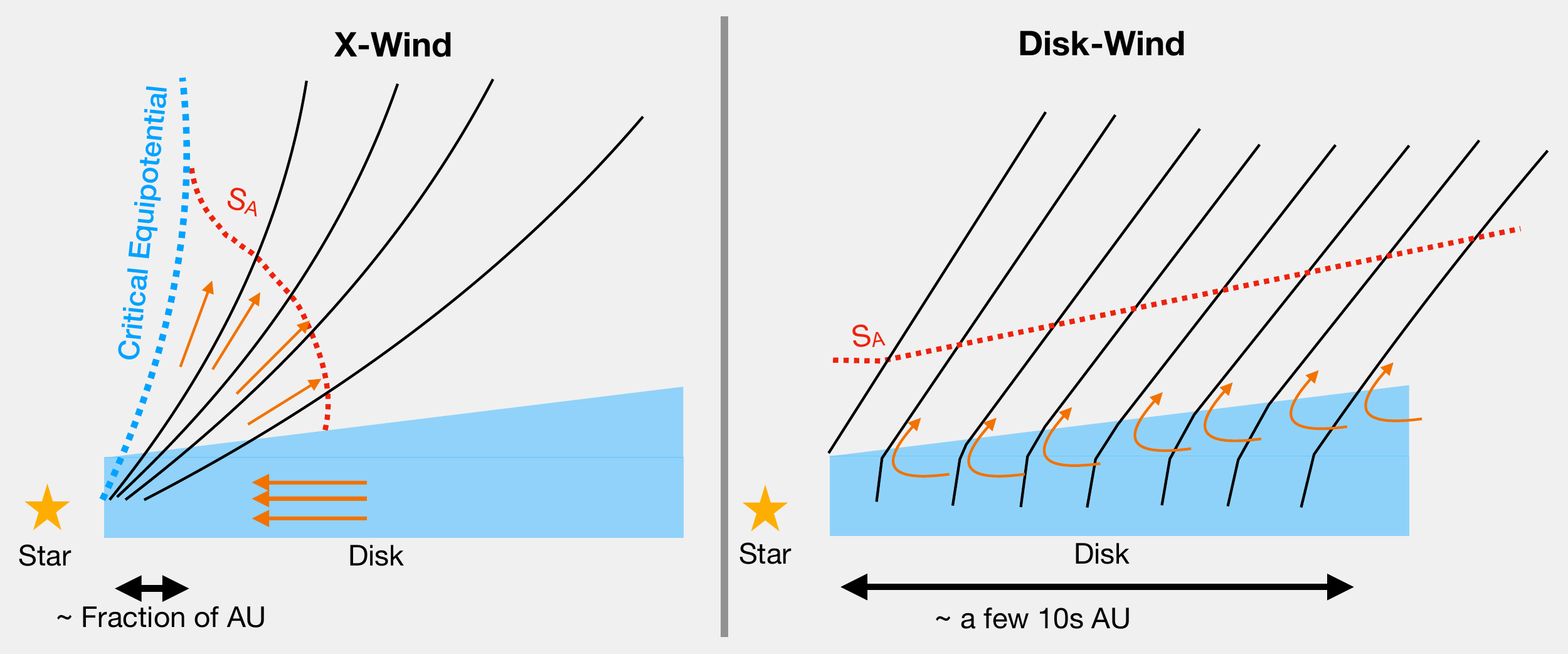}
        \caption{A simplified schematic illustration of magnetically driven winds from protostellar disks. (\textbf{Left}): X-wind launched from a narrow region near the inner disk edge. (\textbf{Right}): Disk wind launched from a wider range of disk radii. In both panels, black lines represent magnetic field lines guiding the outflow, orange arrows indicate the direction of outflowing plasma along the field lines, and the red curve (S$_A$) denotes the Alfv\'en surface, where the flow speed equals the Alfv\'en speed.}  
    \label{fig:schematic_x-wind_disk-wind}
\end{figure}

Several theoretical models have been proposed to explain the launching of jets and winds from protostellar systems, broadly classified into three main categories: X-winds, disk winds, and stellar winds.  The \textit{X-wind} model posits that jets are launched from a narrow region near the disk truncation radius, close to the corotation point with the star (typically fraction of one AU), where magnetic fields efficiently extract angular momentum and drive collimated outflows \citep{1994ApJ...429..781S,2007prpl.conf..261S}, as illustrated in the schematic diagram shown in Figure~\ref{fig:schematic_x-wind_disk-wind}. In contrast, disk wind models propose that winds originate from a broader range of disk radii (typically 0.1 to a few 10 s of AU), with magnetic field lines anchored over an extended region of the disk surface launching material along open field lines via magneto-centrifugal acceleration \citep{1982MNRAS.199..883B,2006A&A...453..785F,2007prpl.conf..277P}, as demonstrated in Figure \ref{fig:schematic_x-wind_disk-wind}. Disk winds generally produce wide-angle outflows and can efficiently remove angular momentum from the disk, supporting sustained accretion. In addition, thermally driven photoevaporative winds, launched from the disk surface by stellar UV and X-ray irradiation, provide another pathway for disk mass loss and can complement MHD-driven winds  (e.g., \citep[]{2021ApJ...913..122R,2025ApJ...986..161H}). It is important to note that recent studies suggest that both X-winds and MHD disk winds can drive highly collimated jets as well as broader, less-collimated outflows, implying that the traditional separation between jets and winds may be overly simplistic  (e.g., \citep[]{2020ApJ...905..116S}). A third scenario involves stellar winds, driven by magnetic pressure or thermal gradients directly from the protostar itself \citep{2005ApJ...632L.135M}.

While stellar winds alone are unlikely to explain the highly collimated jets observed in young protostars, they may contribute to inner outflow regions. Recent high-resolution observations suggest that both X-winds and disk winds may operate simultaneously or sequentially in some systems, with disk winds responsible for wide-angle molecular outflows and X-winds accounting for high-velocity, collimated jets \citep{2014prpl.conf..451F, 2020A&A...640A..82T}. However, distinguishing between these models observationally remains challenging, as signatures often overlap.

\section{Future Works}

Future research on protostellar jets and outflows will greatly benefit from the synergy between infrared and (sub)millimeter observations, particularly combining the capabilities of the James Webb Space Telescope (JWST) and the Atacama Large Millimeter/submillimeter Array (ALMA). JWST provides unprecedented sensitivity and spatial resolution in the infrared, ideal for probing warm molecular and atomic gas, as well as shock-excited regions along jets and outflow cavities \citep{2022ApJ...936L..14P}. ALMA, in turn, offers high-resolution imaging of cold molecular gas, shocked gas and dust structures at (sub)millimeter wavelengths, allowing detailed studies of jet launching regions, disk kinematics, and entrained outflows \citep{2016Natur.540..406B,2020A&A...640A..82T}. Together, these facilities enable multi-wavelength studies that can trace jets across a wide range of temperatures and physical conditions, offering new insights into the interplay between accretion, ejection, and envelope clearing. Another promising avenue involves exploring the impact of jet feedback on planet formation. Jets and outflows can regulate disk mass and angular momentum, potentially influencing the formation and migration of planetary cores within the disk \citep{2016ApJ...821...80B,2021ApJ...915...90N}. Incorporating jet-driven disk evolution into planet formation models is an emerging field, with future observational and theoretical efforts likely to clarify how early protostellar feedback shapes the initial conditions for planet formation.

\section{Summary and Conclusions}

Jets and outflows play a central role in the evolution of protostars, acting as key agents in low-mass star formation by regulating the transport of mass, momentum, and angular momentum, and by influencing the surrounding environment.  Traced by molecular, ionized, and atomic species at (sub)millimeter and infrared wavelengths, these flows provide crucial insights into the accretion--ejection connection and the early stages of protostellar evolution.

Jets are typically highly collimated, high-velocity components launched from the innermost disk regions, while broader, slower disk winds emerge from larger disk radii. Both components play complementary roles in regulating disk evolution and accretion.

Recent advances from high-resolution facilities such as the Atacama Large Millimeter/submillimeter Array (ALMA) and the James Webb Space Telescope (JWST) have dramatically enhanced our ability to resolve the physical and chemical structures of jets and outflows across multiple scales and wavelengths. ALMA has provided unprecedented details on molecular gas kinematics, jet launching regions, and disk winds, while JWST has unveiled deeply embedded infrared counterparts and shock-excited features in outflow cavities and jets.

Despite significant progress, challenges remain in fully disentangling different launching mechanisms, including X-winds, disk winds, and stellar winds. No single model can yet explain the diversity of observed morphologies, kinematics, and chemical properties. Future coordinated, multi-wavelength studies with ALMA, JWST, and upcoming facilities promise to further constrain jet launching physics and their role in disk evolution and planet formation. Although this review is observationally oriented, it might be worth mentioning that models must also incorporate more realistic physics and chemistry and reach finer spatial and temporal resolutions, in concert with modern observations with state-of-the-art facilities.

Finally, jet-driven feedback is emerging as a critical factor in shaping planet-forming environments. Incorporating jet and outflow physics into planet formation models will be essential for developing a comprehensive picture of how planetary systems originate within evolving protostellar disks.

%%%%%%%%%%%%%%%%%%%%%%%%%%%%%%%%%%%%%%%%%%%%%%%%%%%%%%%%%%%%%%%%%%%%%%%%%%%%%%%%%%%%%%%%%%%%%

%%%%%%%%%%%%%%%%%%%%%%%%%%%%%%%%%%%%%%%%%%

%%%%%%%%%%%%%%%%%%%%%%%%%%%%%%%%%%%%%%%%%%
\vspace{6pt} 

%%%%%%%%%%%%%%%%%%%%%%%%%%%%%%%%%%%%%%%%%%
%% optional
%\supplementary{The following supporting information can be downloaded at:  \linksupplementary{s1}, Figure S1: title; Table S1: title; Video S1: title.}


% Only used for preprtints:
% \supplementary{The following supporting information can be downloaded at the website of this paper posted on \href{https://www.preprints.org/}{Preprints.org}.}

%\begin{tabularx}{\textwidth}{lll}
%\toprule
%\textbf{Name} & \textbf{Type} & \textbf{Description} \\
%\midrule
%S1 & Python script (.py) & Script of python source code used in XX \\
%S2 & Text (.txt) & Script of modelling code used to make Figure X \\
%S3 & Text (.txt) & Raw data from experiment X \\
%S4 & Video (.mp4) & Video demonstrating the hardware in use \\
%\bottomrule
%\end{tabularx}
%}

%%%%%%%%%%%%%%%%%%%%%%%%%%%%%%%%%%%%%%%%%%
%\authorcontributions{S.D. planned the study, carried out the analysis, and wrote the manuscript.}

%\funding{This research is independent and has no funding.}
\noindent
{\bf Funding:} This research is independent and has no funding.

\noindent
{\bf Data Availability:} This review article has utilized the data from the lead authors cited in each Figure. The raw ALMA data is available on the ALMA archive as: ADS/JAO.ALMA\#2018.1.00302.S. The data from  the NASA/ESA/CSA James Webb Space Telescope presented in this paper were obtained from the Mikulski Archive for Space Telescopes (MAST) at the Space Telescope Science Institute. All the {\it JWST} data used in this paper can be found in MAST: {\url{https://doi.org/10.17909/52vn-6191}}.
%

%\acknowledgments
{\bf Acknowledgments:} 
We are thankful to the anonymous reviewers for their valuable suggestions, which helped to improve the overall quality of this review. S.D. thanks Chin-Fei Lee for discussion at various stages of my research since last few years and provide the facilitties for my jet/outflow research. I would like to thank ``ALMASOP project" team for accessing ALMA data and discussion on various stages. S.D. acknowledges the facilities  from the Academia Sinica Institute of Astronomy and Astrophysics, Taiwan.
This paper makes use of the following ALMA data: ADS/JAO.ALMA\#2018.1.00302.S (PI: Tie Liu). ALMA is a partnership of ESO (representing its member states), NSF (USA), and NINS (Japan), together with NRC (Canada), NSC, and ASIAA (Taiwan), as well as KASI (Republic of Korea), in cooperation with the Republic of Chile. The Joint ALMA Observatory is operated by ESO, AUI/NRAO, and NAOJ. 
This work utilizes observations made with the NASA/ESA/CSA James Webb Space Telescope. The data were obtained from the Mikulski Archive for Space Telescopes at the Space Telescope Science Institute, which is operated by the Association of Universities for Research in Astronomy, Inc., under NASA contract NAS 5-03127 for JWST. These observations are associated with JWST GO Cycle 1 (Proposal ID: 1854; PI: Melissa McClure).  All the {\it JWST} data used in this paper can be found in MAST: {\url{https://doi.org/10.17909/52vn-6191}}. During the preparation of this manuscript/study, the author(s) used CASA 5.4 \cite{2007ASPC..376..127M}, Astropy \cite{2013A&A...558A..33A}, matplotlib \cite{2007CSE.....9...90H} for the purposes of plotting and analyses. The authors have reviewed and edited the output and take full responsibility for the content of \mbox{this publication.}

{\bf Conflicts of Interest}
The authors declare no conflicts of interest. This research has no funding.  The funders had no role in the design of the study; in the collection, analyses, or interpretation of data; in the writing of the manuscript; or in the decision to publish the results.

\end{document}